\documentstyle[12pt,psfig]{article}
\begin{document}
\title{How can computer simulations contribute to the understanding of
the dynamics of glasses and glass melts?  \footnote{To appear in {\it
Analysis of Composition and Structure of Glass, and Glass Ceramics}
Eds.: H. Bach and D. Krause (Springer, Berlin, 1998)}}

\author{
Walter Kob and Kurt Binder\\
Institut f\"ur Physik, Johannes Gutenberg-Universit\"at\\
Staudinger Weg 7, D-55099 Mainz, Germany
}

\date{November 30, 1997}

\maketitle

\subsection*{Introduction}
Ever since it became possible to perform computer simulations of
systems with a few hundred particles this method was also used to
investigate the structure and dynamics of supercooled liquids and
glasses. The absence of long range order makes these systems a very
difficult task for any sort of analytical calculations and since
experiments often cannot give all the desired information, be it of
structural or dynamical nature, computer simulations offer a very
convenient way to access such information. The goal of the present
article is to review some exemplary results of investigations in which
the dynamics of supercooled liquids was studied. In this we will focus
on recent studies of the prototypical network glassformer SiO$_2$,
since the lack of space does not permit us to review the large body of
literature which exists on the general subject of computer simulations of
supercooled liquids and glasses. However, the interested reader can
find a more thorough discussion of this subject in the review
articles~\cite{angell81,barrat91,kob95,poole95}.

Before we start to introduce the model we used in our simulations and
to present the results, we briefly discuss what the kind of questions
are that can be addressed by means of computer simulations.  Although
it is of course {\it in principle} possible to make a full scale {\it
ab initio} simulation of any given material, the computational demand
restricts such types of investigations to systems that are relatively
small (less than 100 particles) and to short times (on the order of
10ps). Such system sizes and time scales are for most applications in
the field of glasses too small/short and thus one usually limits oneself
to simulations in which the system is described by a classical force
field. However, it is essentially impossible to devise classical force
fields that reproduce exactly the forces in the real material and in
practice thus one always uses some approximative potentials. Therefore
it is clear that it is most improbable that any simulation will
reproduce {\it all} experimentally determined quantities (density,
bond-angles, viscosity, density of states, etc.). In
Ref.~\cite{pot_test}, e.g.,  it is nicely demonstrated how various
potentials for SiO$_2$ can give very different answers regarding the
absolute numbers of various physical quantities, such as the temperature
at which the model shows a maximum in the density or on the temperature
dependence of the diffusion constant, but the bottom line is that most
reasonable potentials agree at least qualitatively with the
experimental findings.  Thus, if the potential model is sufficiently
realistic it can be expected that such a simulation will reproduce at
least the salient features of the material and that therefore one can
use the simulation to identify trends (e.g., temperature dependence of
the density) or mechanisms (e.g., diffusion mechanism) that are present
in the real material as well.

Despite the fact that usually one does not demand a {\it perfect}
agreement between the results of a simulation and real experiments, it
is often not a simple task to find a potential which gives at least a fair
agreement. An example for this situation is B$_2$O$_3$ for which it
seems to be quite difficult to come up with a good classical
potential~\cite{takada95}. Thus it is quite often the case that the
main obstacle to use computer simulations in order to gain insight into
the properties of supercooled liquids and glasses are not the lack of
computational resources but the shortage of classical force fields that
give a realistic description of the real material. Therefore it would
be most useful if progress would be made in the development of
classical force fields.

\subsection*{Model and Details of the Simulation}

In this section we present the model we used for our SiO$_2$
simulation and discuss some of the technical details of the simulation.
More details can be found in
Refs.~\cite{horbach96,horbach97a,horbach97b,horbach98}.

A few years ago {\it van Beest et al.} introduced a model for SiO$_2$
in which the interactions between the ions are of a pure two-body
type~\cite{beest90}.  Despite the absence of three-body interactions
this model is able to generate the tetrahedral network of silica which
is thought to give the correct local structure of this system.  It was
subsequently shown that this potential gives a good description of the
various crystalline phases of silica~\cite{tse} and of the {\it static}
properties of amorphous SiO$_2$ as well~\cite{vollmayr96,binder_sio}.
Thus it is reasonable to use this potential also to investigate the
{\it dynamical} properties of supercooled silica. The potential energy
between two ions of type $i$ and $j$ ($i,j \in \{$Si,O$\}$), a
distance $r_{ij}$ apart, are given by
\begin{equation}
\phi(r_{ij})=\frac{q_i q_j e^2}{r_{ij}}+A_{ij}e^{-B_{ij}r_{ij}}-
\frac{C_{ij}}{r_{ij}^6}\quad .
\label{eq1}
\end{equation}
The values of the parameters $q_i$, $A_{ij}$, $B_{ij}$ and $C_{ij}$ can
be found in the original publication~\cite{beest90}. The non-Coulombic
part of the potential was truncated and shifted at 5.5\AA. The
simulations were done at constant volume and the density of the system
was fixed to 2.37 g/cm$^3$. The system size was 8016 ions, giving a
volume of the cubic box of (48.37\AA)$^3$. This size is significantly
larger than the ones commonly used in such simulations but necessary in
order to avoid finite size effects which have been shown to be
remarkably pronounced in network glassformers~\cite{horbach96}. The
equations of motion were integrated with a time step of 1.6~fs. The
temperatures investigated were 6100~K, 5200~K, 4700~K, 4300~K, 4000~K,
3760~K, 3580~K, 3400~K, 3250~K, 3100~K, 3000~K, 2900~K and 2750~K. At
all temperatures the length of the runs were longer than the typical
relaxation time of the system and extended at the lowest temperature
over about 20~ns. More details on the simulations can be found in
Ref.~\cite{horbach98}.

\subsection*{Results}
One of the simplest dynamical quantities that can be obtained from a
computer simulation is the diffusion constant $D$ of a tagged particle,
which can be computed from the long time behavior of the mean squared
displacement. Care has to be taken, however, that $D$ is 
determined from a time window in which the mean squared displacement
has indeed reached its asymptotic time dependence, i.e. shows a $t^1$
dependence. Otherwise the computed value of $D$ will be too large.

\begin{figure}[f]
\psfig{file=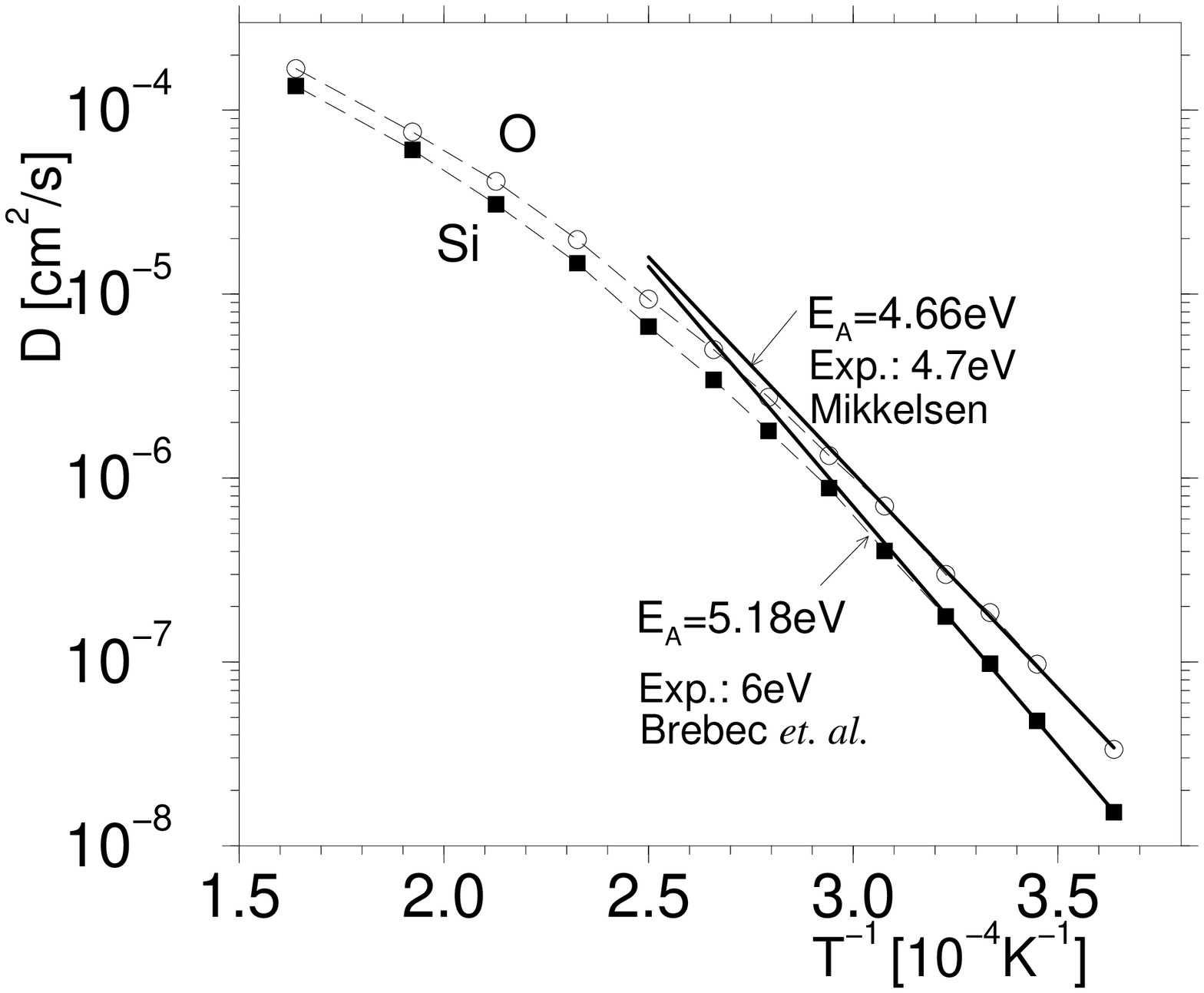,width=13cm,height=9.5cm}
\caption{Temperature dependence of the diffusion constant for
the silicon and oxygen atoms (filled squares and open circles,
respectively). The straight lines are fits to the low temperature data
with an Arrhenius law giving the stated activation energies. The
experimental values for the activation energies (from
Refs.~\protect{\cite{mikkelsen84}} and~\protect{\cite{brebec80}}) are
also given. Adapted from Ref.~\protect{\cite{horbach97a}}. }
\label{fig1}
\end{figure}

In Fig.~\ref{fig1} we show the temperature dependence of
$D_{{\rm Si}}$ and $D_{{\rm O}}$, the self diffusion constant for the
silicon and oxygen atoms, respectively.  We see that for all
temperatures investigated $D_{{\rm O}}$ is larger than $D_{{\rm Si}}$,
which is to be expected since the oxygen atoms are, on average, bound
to the network by only two covalent bonds, whereas the silicon atoms
are bound by four bonds. For low temperatures the diffusion constants
show the Arrhenius temperature dependence well known from experiments.
The two straight lines are an Arrhenius fit to the low temperature data
with activation energies of 4.66eV and 5.18eV for the oxygen and
silicon atoms, respectively.  These values agree well with the
experimental values determined by {\it Mikkelsen}~\cite{mikkelsen84}
and {\it Br\'ebec et al.}~\cite{brebec80}, see figure, and thus we
conclude that the used model for SiO$_2$ is indeed able to give a quite
realistic description of the dynamics of supercooled silica. (We note
that there is some evidence that for the model of van Beest {\it et
al.} the activation energy for silicon might, at low temperatures, be
even a bit larger than the value obtained from the presented
fit~\cite{horbach98}.) From the figure we also recognize that at higher
temperatures the diffusion constants deviate from the Arrhenius
behavior observed at low temperatures in that they are smaller than
expected from an extrapolation with the Arrhenius law. So far this
behavior has not been observed in real experiments, since the
temperatures accessed so far in the experiments are still a bit lower
than the ones for which the non-Arrhenius behavior is observed.  But of
course we cannot exclude the possibility that the observed
non-Arrhenius behavior is an artifact of the used potential with no
counterpart in the real material. On the other hand, the investigation
of {\it Hemmati et al.} has shown that many silica models show such an
non-Arrhenius behavior, although at different temperatures, thus giving
support to the expectation that also real silica will show this
behavior at sufficiently high temperatures.

In order to gain some insight why at high temperatures the temperature
dependence of the diffusion constants seems to be different from the
one at low temperatures, we can study the dynamics of the particles in
more detail. This can be done with the help of the self part of the van
Hove correlation function which is defined as~\cite{hansenmcdonald86}

\begin{equation}
G_s^{(\alpha)}(r,t)=\frac{1}{N_{\alpha}}\left\langle\sum_{i=1}^{N_{\alpha}}
\delta(r-|\vec{r}_i(t)-\vec{r}_i(0)|)\right\rangle\qquad , \alpha \in
\{\mbox{Si,O}\}\quad.
\label{eq2}
\end{equation}
Here $\langle.\rangle$ stands for the thermal average.  Thus
$G_s^{(\alpha)}(r,t)$ is the probability that a particle of type
$\alpha$ has moved within time $t$ a distance $r$.  In Fig.~\ref{fig2}
we show $4\pi r^2 G_s^{({\rm O})}(r,t)$ versus $r$ for different
times.  From these figures we recognize that at high temperatures,
Fig.~\ref{fig2}a, the function is a single peaked function with a
maximum that is moving to larger values of $r$ when time increases.
With decreasing temperature the speed with which the peak moves to
larger distances decreases, but qualitatively the shape of the curves
is similar to the ones shown in Fig.~\ref{fig2}a. This behavior is very
similar to the one observed for {\it fragile} 
glassformers~\cite{barrat91,kob95}, where at low temperatures the dynamics
of the particles is described very well by the so-called mode-coupling
theory~\cite{gotze92,kob97}, which predicts such a time and space
dependence of $G_s(r,t)$. If the temperature is lowered further, say
below 3500~K, this qualitative picture changes, in that for
intermediate times $G_s(r,t)$ shows a double peak structure, see the
curve for $t=4.5$~ns in Fig.~\ref{fig2}b. This secondary maximum,
marked by an arrow, is usually associated with the presence of
so-called hopping processes, i.e. types of movements in which the
particle jumps abruptly to a site in the network which is about one
interparticle distance away from the starting place~\cite{barrat91}.
These sort of processes are usually activated, since the jumping
particle has to overcome a barrier formed by its surrounding particles
(this is in contrast to the motion at higher temperatures, where the
dynamics of the particles is smoother, i.e. where no jumps occur, see
Fig.~\ref{fig2}a). Thus the study of the van Hove correlation function
shows us that the dynamics of the particles changes from a smooth
movement to a jump-like motion when the temperature is lowered. It is
this change in the microscopic dynamics that gives rise to the change
in the temperature dependence of the diffusion constants.

A further important transport quantity is the shear viscosity $\eta$
which can also be measured in real experiments. Unfortunately the
calculation of $\eta$ is not a simple task in computer simulations,
since it is necessary to average this quantity over quite a few (at
least 10) $\alpha$-relaxation times (defined below) before $\eta$ is
determined with a reasonable accuracy. Therefore we were able to
measure $\eta$ only for temperatures larger than 3000~K. The
temperature dependence of $\eta$ is shown in Fig.~\ref{fig3} in an
Arrhenius plot. Despite the noise in the data we can recognize that
this temperature dependence is not Arrhenius like thus is similar to
the temperature dependence of the diffusion constant. Also included in
the figure are the experimental data points of the measurements of {\it
Urbain et al.}~\cite{urbain82}. We see that a reasonable extrapolation
of our data to lower temperature will underestimate the viscosity at
the temperatures at which the experimental values are available, which
shows that with respect to this quantity the potential of van Beest
{\it et al.} is perhaps not quite reliable.  (However, it should also
be kept in mind that the experimental data points might be subject to
an appreciable systematic error.)

Since experimentally the viscosity is simpler to measure than the
diffusion constant, it is often common to calculate the latter from
the former by means of the so-called Eyring equation (which is
essentially identical to the Stokes-Einstein equation), i.e.
\begin{equation}
D=k_BT/\eta\lambda .
\label{eq3}
\end{equation}
Here $k_B$ is Boltzmann's constant and the constant $\lambda$ (having
the dimension of length) is related to the size of an elementary
diffusion step and is usually determined from measurements at lower
temperatures. Having measured both, $D$ and $\eta$, we can now check
whether the combination $k_BT/\eta D$ is indeed constant. This
quantity is shown in the inset of Fig.~\ref{fig3} for the silicon as
well as the oxygen diffusion constant. From this plot we clearly see
that in the temperature range investigated the assumption that
$\lambda$ is constant is a poor approximation. It is, however,
interesting that at the lowest temperatures the value of $\lambda$ for
the oxygen atoms is about 5\AA, which is not too different from 2.8\AA,
the values assumed by {\it Poe et al.}~\cite{poe97} at lower
temperatures.

The fact that the atoms of the system form a three dimensional open
network implies that the bonds between adjacent atoms are broken if an
atom diffuses. Therefore it is reasonable to look into the time
dependence of these bonds. For this we define the quantity $P_b(t)$
which is the probability that a bond which is present at time $t=0$ is
also present at time $t$.  (Note that we call a silicon and oxygen atom
bonded if their distance is smaller than 2.35\AA, the location of the
minimum between the nearest and second nearest neighbor peak in the
Si-O radial distribution function. This location is essentially
independent of temperature.)

In Fig.~\ref{fig4} we show the time dependence of $P_b$ for all
temperatures investigated. From this figure we recognize that at low
temperatures the curves do not decay to zero within the time span of
our simulation. From our investigation of the intermediate scattering
function, discussed below, we know however, that this correlation
function, which measures the structural relaxation of the system, does
decay within the time span of our simulation. Thus we come to the
conclusion that for the structural relaxation to occur it is not
necessary that {\it all} Si--O bonds are broken, but only about 80\% of
them. This number is quite reasonable since it can be argued that an
oxygen atom can make a relaxational movement if one or two of its bonds
are broken and that a silicon atom can move if three of its bonds are
broken, i.e. about 75\% of the average number of bonds.

From Fig.~\ref{fig4} we also see that the shape of the curves is
essentially independent of temperature (a {\it small} temperature
dependence is observed for times at which $P_b$ is still close to its
initial value~\cite{horbach98}). Thus we come to the conclusion that
the mechanism that leads to the breaking of the bonds is also
essentially independent of temperature. Thus it is reasonable to define
the mean lifetime $\tau_b$ of a bond by requiring that
$P_b(\tau_b)=e^{-1}$. This lifetime can now be compared with the
typical time for the diffusion. For this we plot in Fig.~\ref{fig5} the
products $D_{{\rm Si}}\tau_b$ and $D_{{\rm O}}\tau_b$ as a function of
temperature. We recognize from this figure that this product is
essentially constant for the oxygen atoms, but decreases for the
silicon atoms. Thus we conclude that the breaking of a bond is related
to the diffusive movement of the oxygen atom, which is very reasonable,
since this atom becomes mobile after the breaking of one of its bonds
to its two nearest neighbors silicon atoms. For silicon atoms the
situation is different: Even if one bond to its neighboring oxygen
atoms is broken it is still tightly bound to the network by the
remaining oxygen neighbors and thus cannot start to diffuse around.
Therefore the quantity $\tau_b$ is not related to the diffusion
constant of the silicon atoms and hence $\tau_b D_{{\rm Si}}$ is not
constant.

As mentioned above, the structural relaxation of the system can be
studied very well by means of $F_s(q,t)$, the (incoherent)
intermediate scattering function for wave-vector $q$. A further reason
why this function is important is that it can be measured in neutron
and light scattering experiments. It is defined as~\cite{hansenmcdonald86}
\begin{equation}
F_s^{(\alpha)}(q,t)=\frac{1}{N_{\alpha}}\left\langle \sum_{j=1}^{N_\alpha}
\exp\left(iq\cdot(\vec{r}_j(t)-\vec{r}_j(0)\right)\right\rangle 
,\qquad \alpha \in \{\mbox{Si,O}\}\quad.
\label{eq4}
\end{equation}
In Fig.~\ref{fig6}a we show the time dependence of $F_s^{({\rm
O})}(q,t)$ for the wavevector $q=1.7$\AA$^{-1}$ which corresponds to
the location of the first sharp diffraction peak in the static
structure factor. We see that at high temperatures the curves decay
quickly to zero with a time dependence which is approximated well by an
exponential function. When the temperature is lowered a shoulder starts
to form at around 0.4~ps which, upon lowering the temperature further,
develops into a plateau. The time range during which the correlation
function is close to the plateau is usually called the
$\beta$-relaxation regime, whereas the time range during which the
correlator starts to fall below the plateau and to zero is called the
$\alpha$-relaxation regime. The physical significance of this plateau
is that in the time range where it is observed the particles rattle
around in the cage formed by the neighboring particles and only for
larger times this cage starts to break up, the particles begin to
diffuse and thus the correlation function starts to decay to zero.

From Fig.~\ref{fig6}a we also see that {\it at low} temperatures the
shape of the curves seems, {\it in the $\alpha$-relaxation regime}, to
be independent of temperature. To check whether this is indeed the case
we define the $\alpha$-relaxation time $\tau(q)$ by requiring that
$F_s(q,\tau(q))=e^{-1}$. Thus, if the shape of the curves are in the
$\alpha$-relaxation regime indeed independent of temperature, a plot of
the curves versus the {\it rescaled} time $t/\tau(q)$ should give a
master curve. (If the correlation functions show such a scaling
behavior one also says that they obey the {\it time-temperature
superposition principle}.) That this is indeed the case is demonstrated
in Fig.~\ref{fig6}b, where we plot the same correlation function as
shown in Fig.~\ref{fig6}a versus $t/\tau(q)$. We see that we find
indeed that the curves for the different temperatures fall, for low
temperatures, reasonably well onto a master curve. Thus we can conclude
that the slowing down of the relaxation dynamics with decreasing
temperature is {\it not} due to the fact that the decay of the
correlation functions at long times becomes slower with decreasing
temperature but rather that the time until the correlation functions
start to deviate appreciably from the plateau, i.e. the time for the
breaking up of a cage, increases strongly with decreasing temperature.

From Fig.~\ref{fig6} one also sees that at times around 0.2-0.5~ps
the correlation functions for low temperatures show a noticeable dip.
Such a feature is {\it not} observed in hard-sphere like systems, such
as Lennard-Jones or soft spheres~\cite{barrat91,kob95,kob95b}, which
are fragile glassformers, and thus seem to be a particularity of
strong glassformers.  This dip is related to the so-called boson peak,
a feature which has been observed in various experiments, such as
inelastic neutron scattering, Raman scattering and more recently also
in inelastic x-ray scattering. The reason for the presence of this peak
is so far a matter of dispute~\cite{boson_peak} since the experiments are
not able to provide the information necessary to decide between the
different theoretical models. Thus this type of question might very
well be one which will ultimately be decided by means of specific
types of computer simulations, since those will allow to access the
necessary information, and some efforts in this direction have indeed
been made~\cite{taraskin97a}.

The last quantities we discuss in this article are the dispersion
relations of the different modes. Within a computer simulation it is
relatively simple to calculate the wave- and frequency dependence of the
(collective) dynamic structure factors 

\begin{equation}
S_{\alpha\beta}(q,\omega)=\frac{1}{2\pi}\int_{-\infty}^{\infty}
F_{\alpha\beta}(q,t)\exp(i\omega t)dt\quad\alpha,\beta\in
\{\mbox{Si,O}\}\quad,
\label{eq5}
\end{equation}
with the coherent intermediate scattering functions
\begin{equation}
F_{\alpha\beta}(q,t)=\frac{N_{{\rm Si}}+N_{{\rm O}}}{N_{\alpha}N_{\beta}}
\int d\vec{r} \exp(-i\vec{q}\cdot\vec{r}) \left\langle
\sum_{k=1}^{N_{\alpha}}\sum_{j=1}^{N_{\beta}}
\delta(r-|\vec{r}_j(t)-\vec{r}_k(0)|)
\right\rangle ,
\label{eq6}
\end{equation}
since all the necessary information to compute this quantity is readily
accessible. The dynamic structure factors are interesting for at least
two reasons: Firstly they can be also measured in experiments, although
only in a relatively limited range of frequency and wave-vectors, and
secondly they allow insight into the nature of the excitations present
in the system, such as the acoustic modes and the optical modes. The
fact that we can compare the results of $S_{\alpha\beta}(q,\omega)$
from a simulation with real experimental data gives us the possibility
to test whether the silica model used is able to reproduce this type of
dynamic observable in a satisfactory way. In the case where such a
comparison is satisfactory we then can thus use the results from the
simulation to gain information on $S_{\alpha\beta}(q,\omega)$ in those
$q$ and $\omega$ ranges which are not accessible to real experiments.
Such comparisons have indeed been made and it was shown that the
present models for silica agree quite reasonably with
experiments~\cite{horbach98,taraskin97a}.

Once the spectra are obtained as a function of frequency the location 
of the various peaks can be read off and thus the dispersion relation
of the different modes determined. (Note that for the investigation
of the $q$-dependence of the {\it transversal} modes it is necessary to 
compute the transversal current-current correlation function
$C_t(q,\omega)$, which can also be expressed simply by means of the
positions and velocities of the particles~\cite{hansenmcdonald86}.)

In Fig.~\ref{fig7} we show the so obtained dispersion relations for the
different modes at 2900~K~\cite{horbach97b,horbach98}.  The data for
the Si--Si and the O--O correlation are shown in filled and open
symbols, respectively. Starting at small wavevectors, various branches
can be observed for each species: Two optical ones in the transverse
and longitudinal modes, denoted by LO1, LO2, TO1 and TO2, and a
longitudinal and transversal acoustic mode (LA and TA, respectively). In
addition to these modes we find at wavevectors larger than
0.22\AA$^{-1}$ an {\it additional} peak in the dynamic structure factor
at around 2THz (see Fig.~\ref{fig7}b). From its location at higher
values of $q$, which is always around 2THz and a comparison with
experimental data~\cite{buchenau86,wischnewski97}, we conclude that
this excitation is the so-called boson peak. The wavevector dependence
of the different modes becomes relatively complicated when $q$ is large
and it has been discussed in greater detail in
Refs.~\cite{taraskin97b}. E.g. we see that the longitudinal acoustic
mode for the silicon atoms (open triangles pointing downwards) show a
quasiperiodic $q$-dependence which reflects the fact that the system
has a pseudo-Brillouin zone at around 1.4\AA$^{-1}$~\cite{taraskin97b}.
Presently we do not want to dwell further on the discussion of the
various curves and just point out one last thing. In Fig.~\ref{fig7}b
we have also included two bold solid lines. These present the linear
dispersion behavior of the acoustic modes in the viscoelastic limit.
Note that the slopes of the lines, which is equal to $c_L$ and $c_T$, the
longitudinal and transversal speed of sound, were {\it not} taken as a
fit parameter, but are the experimental speeds of sound at around
1600~K~\cite{wischnewski97}.
From the fact that for small wavevectors our data points lie very well
on (in the case of the longitudinal mode) or at least quite close (in
the case of the transversal mode) to these lines with the experimental
slopes we conclude that our simulations are able to reproduce also this
type of measurements with good accuracy.

\subsection*{Conclusions}

The goal of this review was not to make a comprehensive review of all
the computer simulations of supercooled liquids and glasses. Rather we
tried to present some exemplary results in order to demonstrate how
computer simulations can contribute to our understanding of the
dynamics of supercooled liquids. As already mentioned in the
Introduction, the most valuable use of computer simulations is {\it
presently} not to try to reproduce the dynamical features of the
material of interest to a {\it very} high accuracy. Rather it is
advisable to use rather simple models which are able to reproduce the
{\it essential} features that one is interested in and to make a
careful investigation of the properties of such models. As we have
demonstrated in this article, even such relatively simple models can
give a quite realistic description of reality. Due to their simplicity,
however, it is possible to simulate system sizes and time scales that
would not be accessible with more sophisticated models. Because with
today's computer hardware it is possible to simulate systems that are
reasonably large, on the order of 50\AA, for time spans (10-50~ns) that
are much longer than the microscopic times (1~ps), such simulations can
indeed give valuable information into the dynamics of glass melts and
glasses, such as the lifetime of a bond, the mechanism of diffusion or
the dynamics of the particles on the microscopic scale, quantities that
are experimentally accessible only in a limited way, or not at all.
Therefore these types of computer simulations are a valuable
complement to experimental investigations to understand the dynamics
of such systems.

\vspace*{2mm}\par
\noindent
Acknowledgements: We thank J. Horbach who provided us with the results
presented here and for a critical reading of the manuscript.
Furthermore we benefited from discussions with C.A. Angell, U.
Buchenau, M. Hemmati, G. Ruocco, F. Sciortino and K. Vollmayr on
various aspects of this work.  We also thank U. Fotheringham, D.
Krause, and W. Pannhorst from SCHOTT Glaswerke for suggesting us to
investigate network glassformers, and the Deutsche
Forschungsgemeinschaft (DFG, grant No.  SFB 262/D1) and the
Bundesministerium f\"ur Bildung, Forschung, Wissenschaft und
Technologie (BMBF, grant No. 03N8008C) for financial support for this
research. Last not least we thank the computing centers in J\"ulich and
Stuttgart for permitting us to use their CRAY T3E without which these
simulations would not have been possible.

\clearpage
\newpage

\begin{figure}[f]
\psfig{file=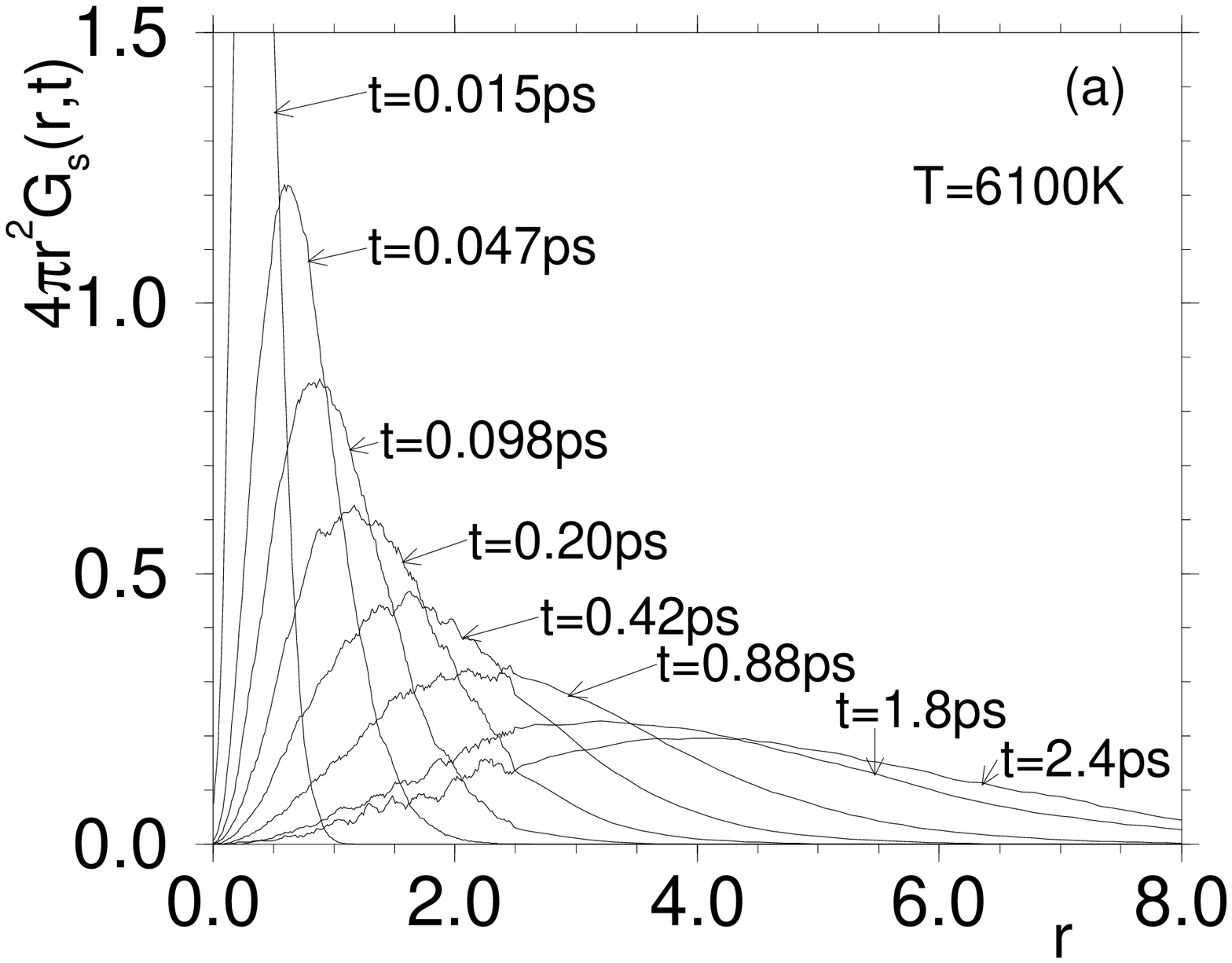,width=13cm,height=9.5cm}
\end{figure}
\begin{figure}[f]
\psfig{file=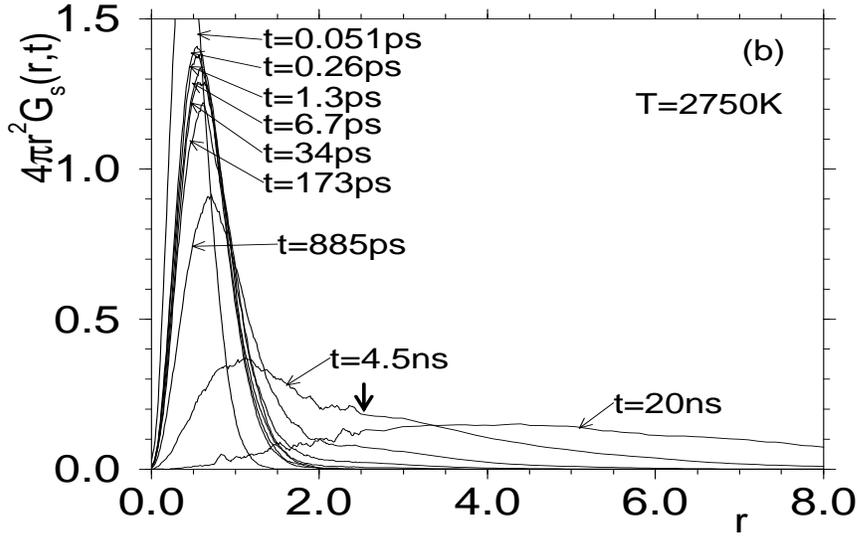,width=12cm,height=8.0cm}
\caption{The self part of the van Hove correlation function for the
oxygen atoms versus distance $r$ for times that are evenly spaced on a
logarithmic time axis (see labels). a) $T=6100$~K, b) $T=2750$~K. The
arrow marks the location of the secondary peak. From
Ref.~\protect{\cite{horbach98}}. }
\label{fig2}
\end{figure}
\begin{figure}[f]
\psfig{file=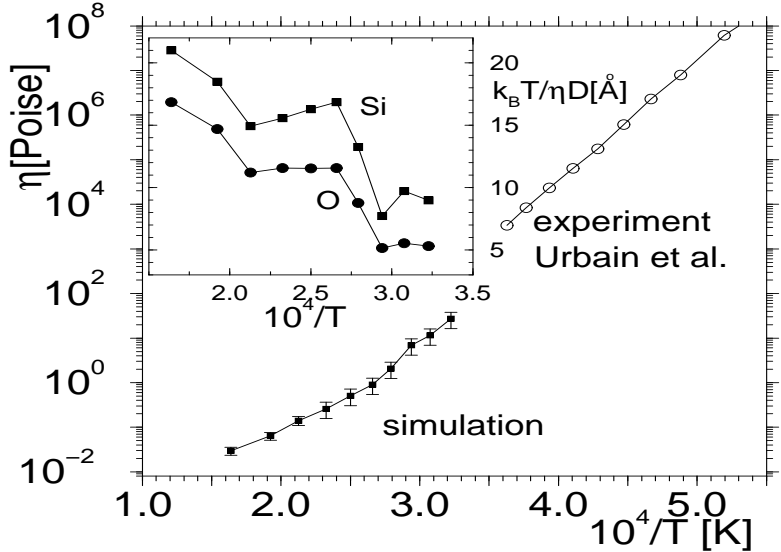,width=12cm,height=8.0cm}
\caption{Main figure: Temperature dependence of the viscosity as
determined from the simulation, (filled squares), and from the
experiments of {\it Urbain et al.}, (open circles)
Ref.~\protect{\cite{urbain82}}.  Inset: Temperature dependence of
$k_BT/D\eta$ for the silicon and oxygen diffusion constant (circles and
squares, respectively). From Ref.~\protect{\cite{horbach98}}. }
\label{fig3}
\end{figure}
\begin{figure}[f]
\psfig{file=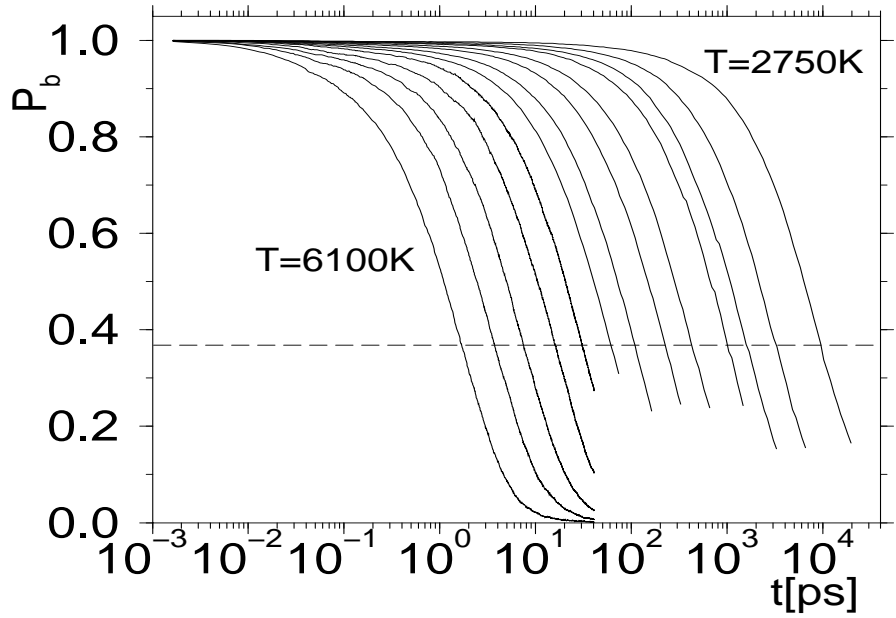,width=13cm,height=9.0cm}
\caption{Time dependence of $P_b(t)$, the probability that a Si--O
bond which was present at time $t=0$ is also present at time $t$, for
all temperatures investigated. The horizontal dashed line at $e^{-1}$
is used to define the lifetime $\tau_b$ of a bond.
From Ref.~\protect{\cite{horbach98}}. }
\label{fig4}
\end{figure}
\begin{figure}[f]
\psfig{file=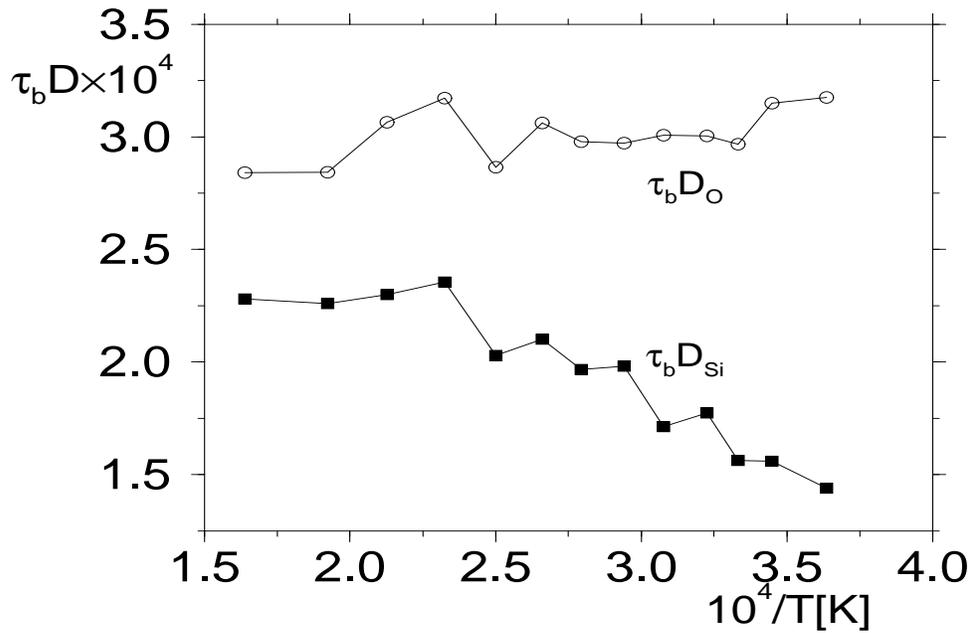,width=13cm,height=9.0cm}
\caption{Temperature dependence of the product between the lifetime
of a Si--O bond and the silicon and oxygen diffusion constant (filled
and open symbols, respectively). From Ref.~\protect{\cite{horbach98}}. }
\label{fig5}
\end{figure}
\begin{figure}[f]
\psfig{file=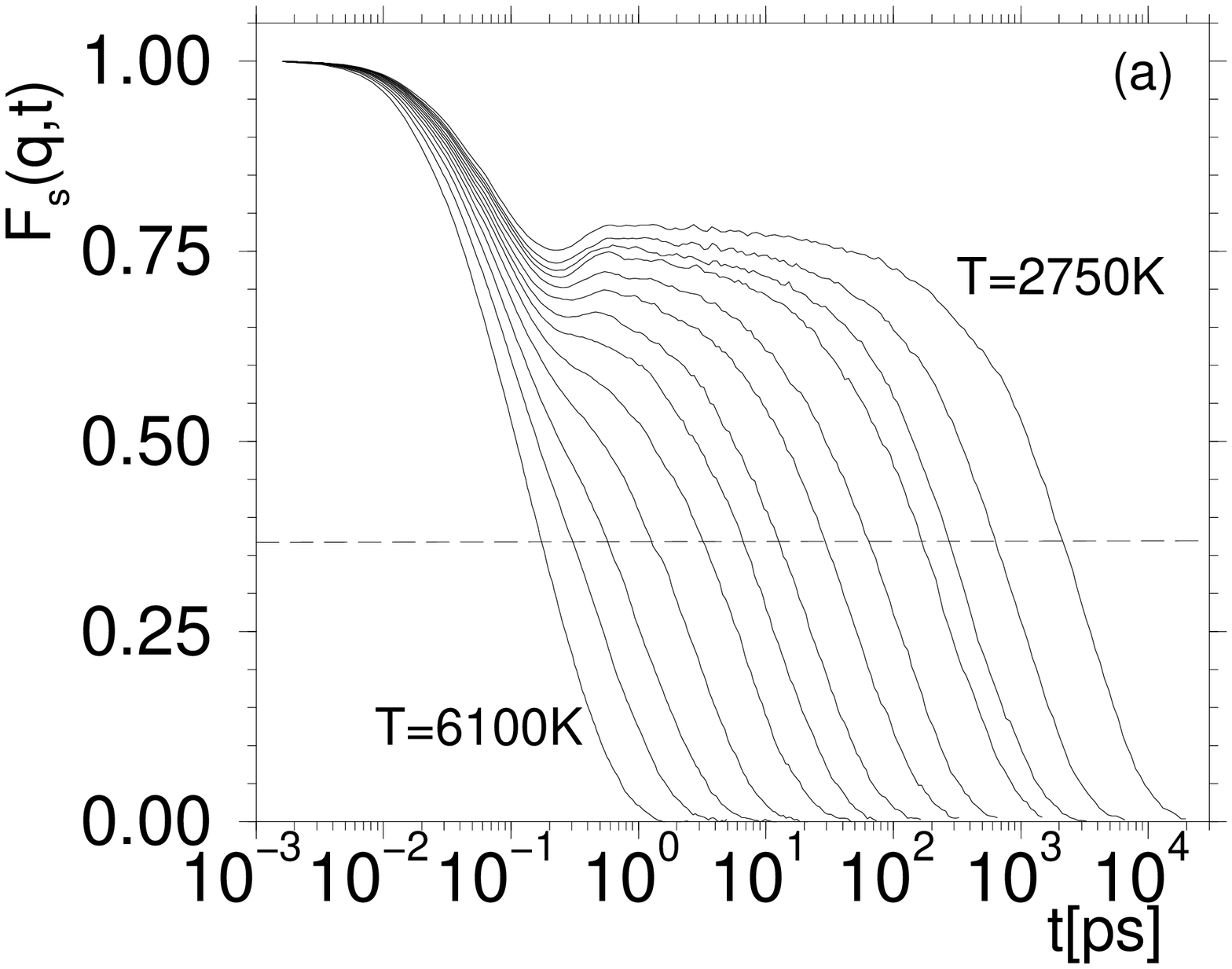,width=13cm,height=9.5cm}
\end{figure}
\begin{figure}[f]
\psfig{file=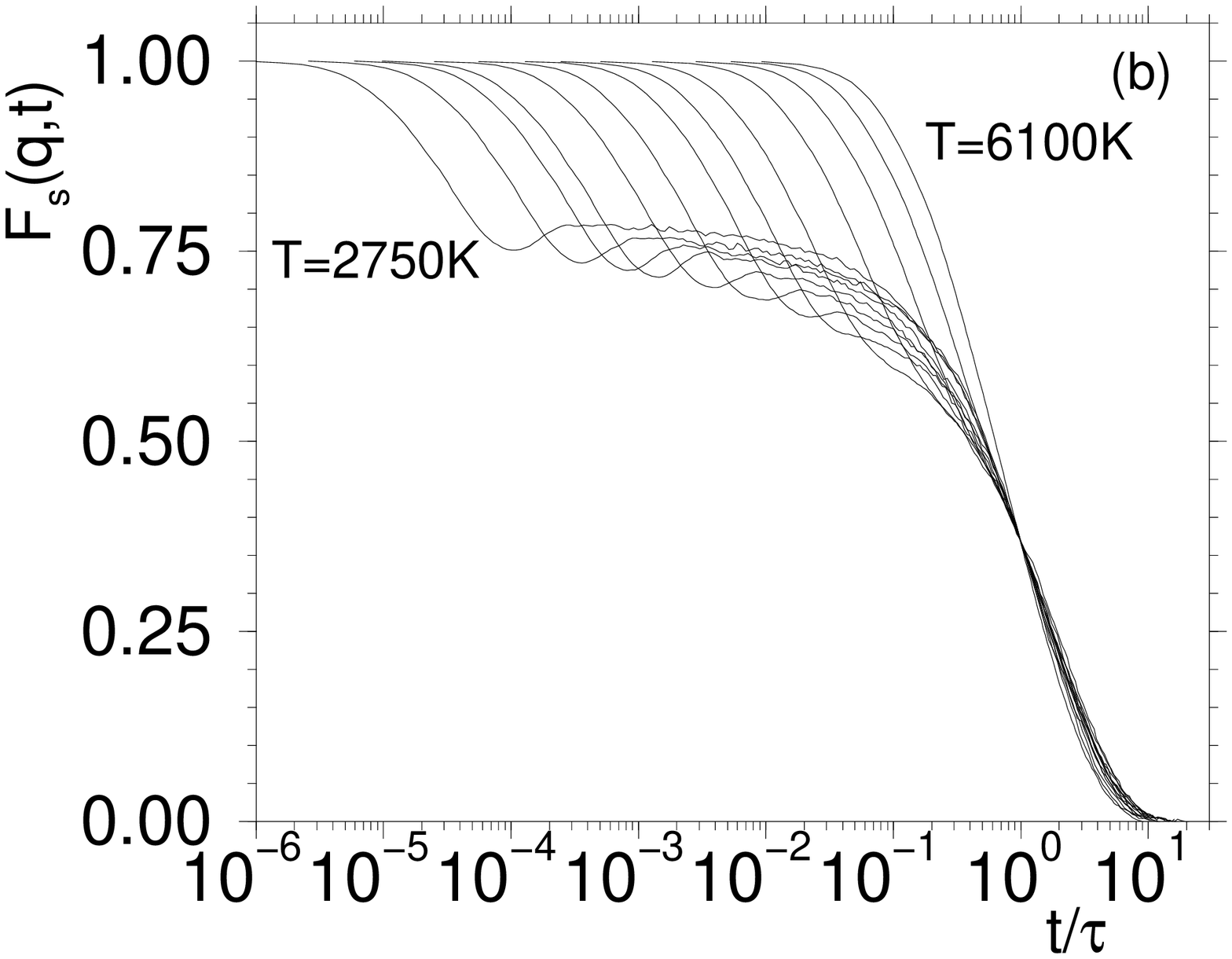,width=13cm,height=9.5cm}
\caption{a) Time dependence of the incoherent intermediate scattering
function $F_s(q,t)$ for the oxygen atoms for all temperatures
investigated. The value of the wavevector $q$ is 1.7\AA$^{-1}$ and
corresponds to the location of the first sharp diffraction peak in the
static
structure factor.  The horizontal dashed line at $e^{-1}$ is used to
define the $\alpha$-relaxation time $\tau$. b) Same data as in a) but
plotted versus the rescaled time $t/\tau(T)$. From
Ref.~\protect{\cite{horbach97a}}.}
\label{fig6}
\end{figure}
\begin{figure}[f]
\psfig{file=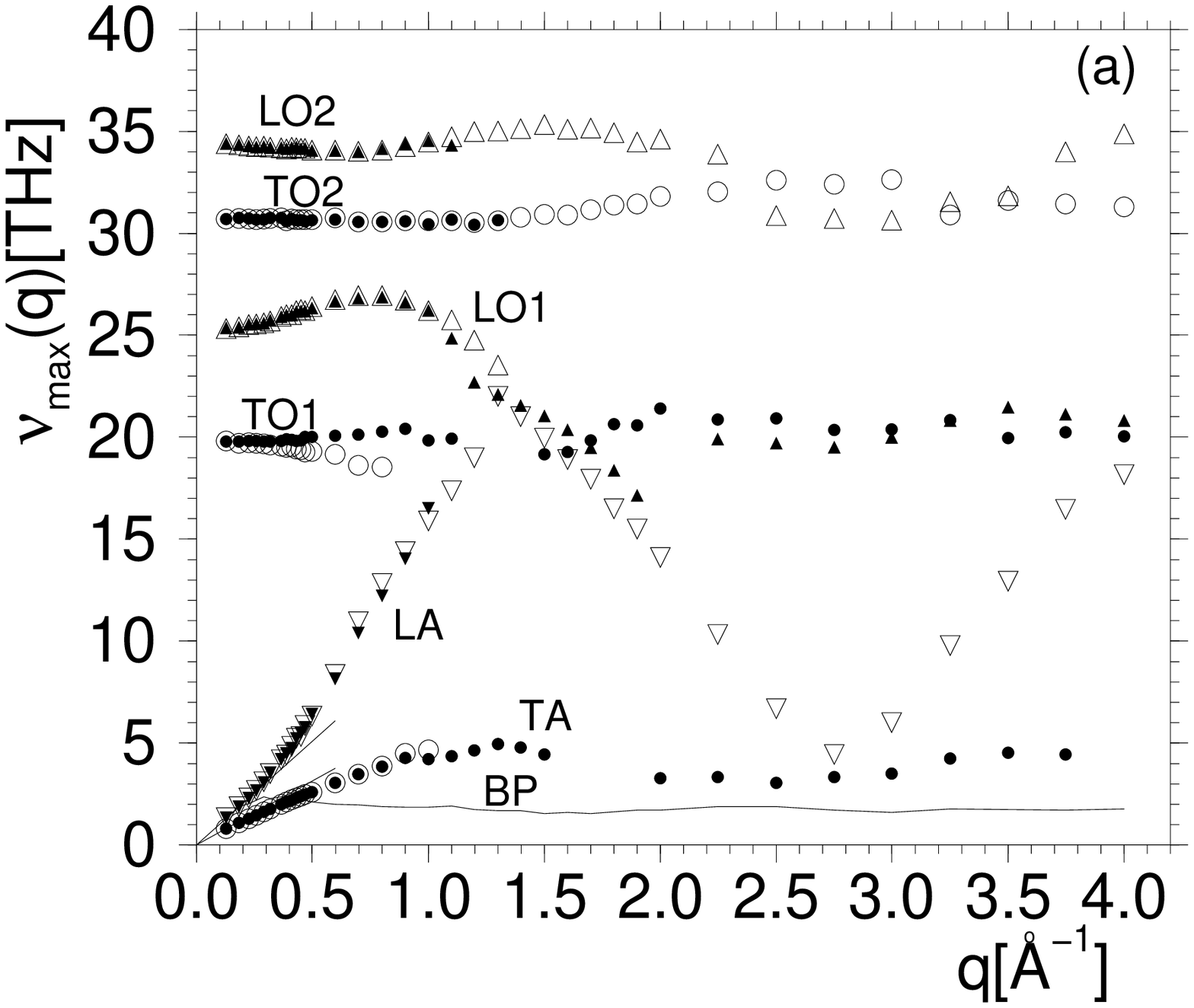,width=13cm,height=9.0cm}
\end{figure}
\begin{figure}[f]
\psfig{file=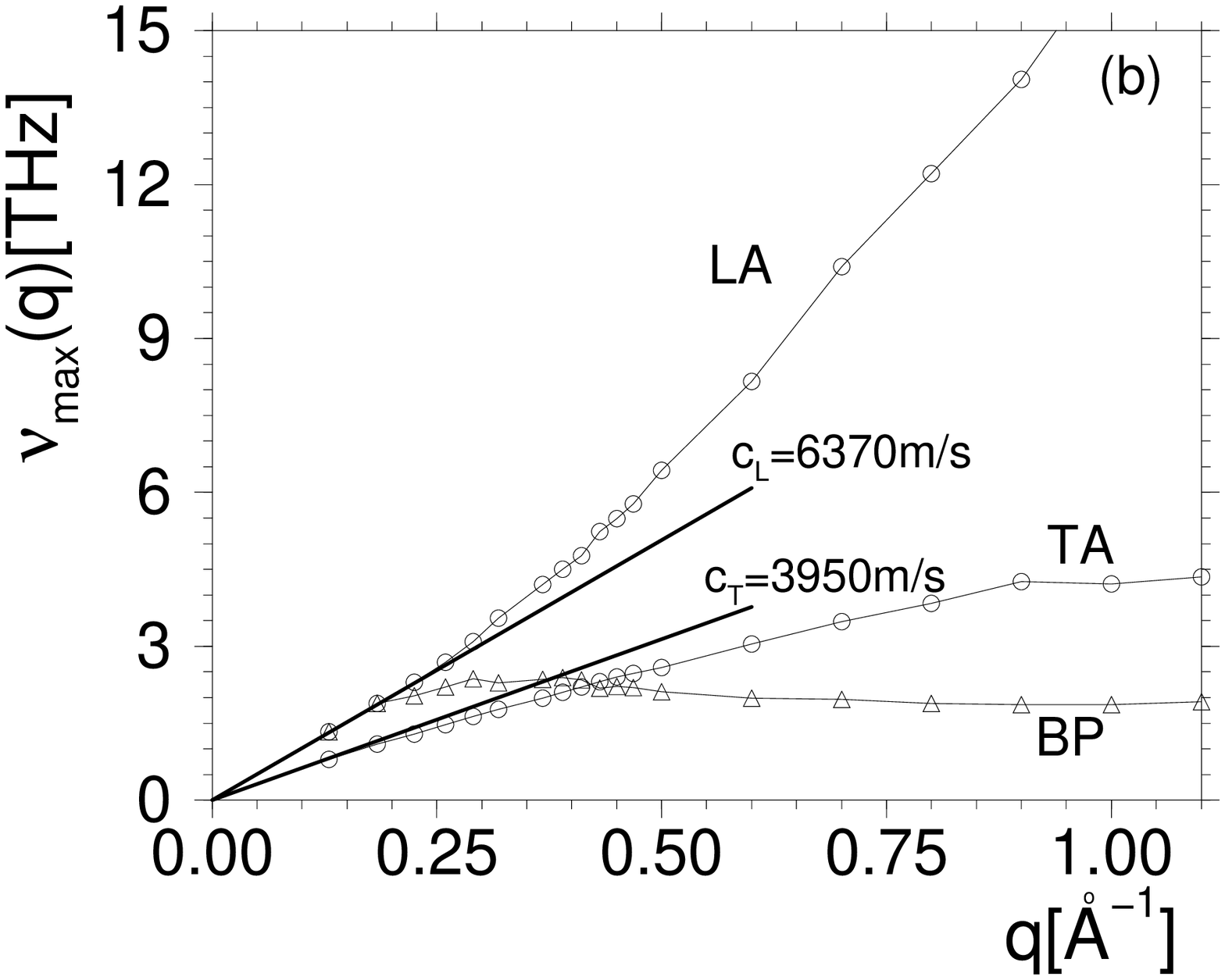,width=13cm,height=9.0cm}
\caption{Dispersion relations for different modes at $T=2900~K$. The
filled and open symbols correspond to the Si--Si and O--O correlation,
respectively.  The first character in the labels (L,T) stands for
longitudinal and transversal and the second character (O,A) for optical
and acoustic. The dispersion relation labeled BP is the one for the
boson peak. a) All wavevectors considered. b) Dispersion relations at
small wavevectors for the O--O correlation. 
The two straight lines are the expected dispersion
relation if the experimental velocities of sound are
assumed~\protect{\cite{wischnewski97}}. Adapted from
Ref.~\protect{\cite{horbach98}}. }
\label{fig7}
\end{figure}
\end{document}